\documentclass[aps,prb,showpacs,amsmath,amssymb,twocolumn,notitlepage]{revtex4-2}

\usepackage{color}
\usepackage{amsmath}
\usepackage{graphicx}
\usepackage{multirow}
\usepackage{float}
\usepackage{bm}

\DeclareMathOperator{\sgn}{sgn} 

\begin{document}

\newcommand{\bn}{{\bm n}}
\newcommand{\bp}{{\bm p}}   
\newcommand{\br}{{\bm r}}
\newcommand{\bk}{{\bm k}}
\newcommand{\bv}{{\bm v}}
\newcommand{\brho}{{\bm{\rho}}}
\newcommand{\bj}{{\bm j}}
\newcommand{\wk}{\omega_{\bf k}}
\newcommand{\nk}{n_{\bf k}}
\newcommand{\eps}{\varepsilon}
\newcommand{\la}{\langle}
\newcommand{\ra}{\rangle}
\newcommand{\be}{\begin{equation}}
\newcommand{\ee}{\end{equation}}
\newcommand{\intl}{\int\limits_{-\infty}^{\infty}}
\newcommand{\dE}{\delta{\cal E}^{ext}}
\newcommand{\SE}{S_{\cal E}^{ext}}
\newcommand{\dsp}{\displaystyle}
\newcommand{\phit}{\varphi_{\tau}}
\newcommand{\p}{\varphi}
\newcommand{\cL}{{\cal L}}
\newcommand{\dphi}{\delta\varphi}
\newcommand{\dbj}{\delta{\bf j}}
\newcommand{\dI}{\delta I}
\newcommand{\dph}{\delta\varphi}
\newcommand{\ua}{\uparrow}
\newcommand{\da}{\downarrow}
\newcommand{\ip}{\{i_{+}\}}
\newcommand{\im}{\{i_{-}\}}
  
\title{Electron-electron scattering and transport properties of spin-orbit coupled electron gas}

\author{K. E.~Nagaev} 
\author{A. A. Manoshin}

\affiliation{Kotelnikov Institute of Radioengineering and Electronics, Mokhovaya 11-7, Moscow 125009, Russia}

\date{\today}

\begin{abstract}
We calculate the electrical and thermal conductivity of a two-dimensional electron gas with strong spin--orbit
coupling in which the scattering is dominated by electron--electron collisions. Despite the apparent absence
of Galilean invariance in the system, the two-particle scattering does not affect the electrical conductivity above
the band-crossing point where both helicity bands are filled. Below the band-crossing point where one helicity band is empty,
switching on the electron--electron scattering leads only to a limited decrease of the electrical conductivity, so that
its high-temperature value is independent of the scattering intensity. In contrast to this, thermal conductivity is not
strongly affected by the spin-orbit coupling and exhibits only a kink as the Fermi level passes through the band-crossing
point.

\end{abstract}

\maketitle

\section{Introduction}

Two-dimensional (2D) systems with spin-orbit coupling (SOC) are key components of spintronics devices \cite{Zutic04}. Apart 
from this, they possess a nontrivial structure of energy bands, which  makes their charge- and heat-transport properties an 
interesting subject of research. So far, the main attention  was focused on 2D SOC systems with purely elastic scattering
\cite{Brosco16,Xiao16,Hutchinson18,Sablikov19,Chen20}. In particular, it was found that the impurity-related resistivity 
exhibits an unconventional dependence on the electron density \cite{Brosco16,Hutchinson18}. Far less attention was given to
the effects of inelastic scattering. Meanwhile it is of interest to find out how the electrical and heat conductivity are
affected by electron--electron collisions. Typically, this scattering affects thermal conductivity but does not contribute
to electrical conductivity in the absence of Umklapp processes, which change the total quasimomentum of colliding electrons
by a reciprocal-lattice vector \cite{Peierls29,Landau36} and take place only if the size of Fermi surface is comparable with
that of the Brillouin zone. However due to the absence of Galilean invariance in systems with SOC, it may give a
nonzero contribution to both of these quantities.

The Rashba spin-orbit coupling \cite{Bychkov84} splits the electron spectrum into the upper and lower helicity bands where
the electron spin is locked to its momentum clockwise or counterclockwise. These bands cross at only one point in the momentum 
plane and the Fermi surface is doubly connected both above and below the corresponding energy. The effect of  
electron--electron scattering on the electric conductance in multiband electron systems was considered in a number of papers 
\cite{Appel78,Murzin98,Hwang03,Pal12-LJP} and it was found to give a contribution to the resistivity proportional to the square of 
temperature $T$. We show that this contribution exists in 2D SOC systems only below the band-crossing point. Moreover, it follows 
the $T^2$ dependence only if the electron--electron scattering is accompanied by a much stronger impurity scattering. As the 
temperature increases, the inelastic contribution saturates and the resistivity tends to a limiting value which is determined only 
by the elastic scattering, in violation of the Matthiessen's rule. In contrast to this, the thermal conductivity limited by 
electron--electron scattering follows the $T^{-1}\ln^{-1}(E_F/T)$  temperature dependence characteristic of 2D systems 
\cite{Lyakhov03} both below and above the band-crossing point, while its dependence on the chemical potential shows a kink at 
this point.

The rest of paper is organized as follows. In Section \ref{sec:general} we describe the model and write down the kinetic equations
for the general case. In Section \ref{sec:electric} we calculate the electrical conductivity in the presence of electron--electron 
and electron--impurity scattering and consider the limiting cases. In Section \ref{sec:thermal}, the thermal conductivity is calculated
in the presence of electron--electron scattering alone, and finally Section \ref{sec:discussion} contains the discussion of the
results. The details of calculations are given in the Appendix.

\section{Model and general equations} \label{sec:general}

\begin{figure}
 \includegraphics[width=0.9\columnwidth]{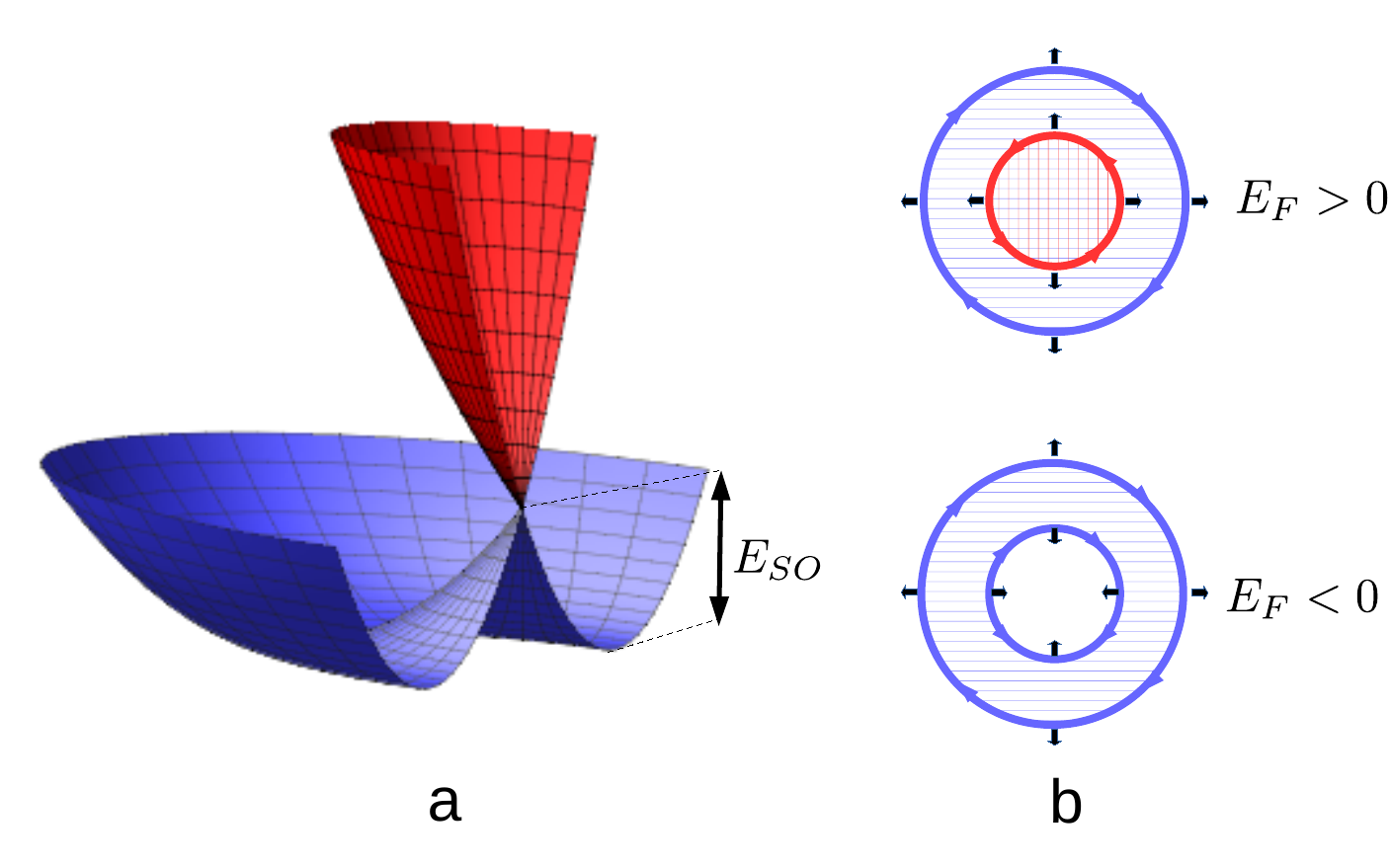}
 \caption{\label{fig:bands} (a) 3D plot of the lower and upper helicity bands touching each other at the Dirac point. (b) The doubly connected Fermi surface above 
 and below the Dirac point. The arrows show the directions of spin at the corresponding Fermi contours, the filled states are hatched with the colors of the corresponding 
 band. Black arrows denote the directions of velocity.}
\end{figure}

We consider a 2D electron gas with  strong Rashba spin-orbit coupling and weak electron--electron and electron--impurity 
interactions, which will be treated as perturbations. If the gas resides in the $xy$ plane, the unperturbed Hamiltonian 
is of the form
\be
 \hat{H} = \frac{\hat p_x^2 + \hat p_y^2}{2m} + \alpha\,(\hat\sigma_x\hat p_y - \hat\sigma_y\hat p_x),
 \label{H}
\ee
where $\alpha$ is the Rashba coupling constant and $\hat{\sigma}_{x,y}$ are the Pauli matrices. This Hamiltonian is easily
diagonalized, and this results in two branches of spectrum
\be
 \eps_{\nu}(\bp) = \frac{p_x^2 + p_y^2}{2m} + \nu\alpha\sqrt{p_x^2 + p_y^2}, \quad \nu=\pm 1,
\label{eps+-}
\ee
which are shown in Fig. \ref{fig:bands}a. These branches give rise to two bands that intersect at only one point at the origin.
While the upper branch $\eps_1(p)$ monotonically increases, the lower branch exhibits a minimum $\eps_{-1}(p_0) = -E_{SO}$, 
where $p_0= m\alpha$ and $E_{SO} = m\alpha^2/2$. This suggests that the Fermi surface of the electron gas is doubly connected at
the Fermi energy $E_F$ both below and above the band-crossing point and consists of two concentric circumferences. However at 
$E_F<0$, the occupied electron states form a ring between these contours and the directions of velocity at them are opposite.
On the contrary, the velocities at both Fermi contours at $E_F>0$ are aligned in the same direction, see Fig. \ref{fig:bands}b.
The corresponding wave functions are spinors 
whose components correspond to the spin projection on the $z$ axis $\sigma=\pm 1/2$
\be
 \psi_{\bp\nu}(\br) 
 = \frac{1}{\sqrt{2}}\,e^{i\bp\br/\hbar} \binom{e^{i\chi_{\bp}/2}}{\nu e^{-i\chi_{\bp}/2}}
 \label{WF}
\ee
with $\chi_{\bp} = \arctan(p_x/p_y)$, so the spin is locked to the momentum and its component perpendicular to $\bp$
is $\pm 1/2$. The sign of this component determines the helicity of the corresponding band.

A response of a fermionic system with weak scattering to slow-varying external fields is conveniently described by the 
standard kinetic equation of the form \cite{Pitaevskii-book}
\be
 \frac{\partial f_{\nu}}{\partial t} 
 + \frac{\partial\eps_{\nu}}{\partial\bp}\,\frac{\partial f_{\nu}}{\partial\br}
 + e{\bm E}\,\frac{\partial f_{\nu}}{\partial\bp}
 = I_{\nu}^{imp} + I_{\nu}^{ee},
 \label{Boltz-1}
\ee
where $I_{\nu}^{imp}$ and $I_{\nu}^{ee}$ describe collisions of electrons with impurities and with each other.
Note that the electron distribution function $f_{\nu}(\bp,\br)$ is the probability of finding an electron in the
state  $|\nu,\bp\ra$ at point $\br$ as in Ref. \cite{Sablikov19} and not the probability of finding there an electron with $z$
projection of spin $S_z=\pm 1/2$, as in many papers on spin transport \cite{Hankiewicz06}. This allows us to write the
collision integrals in the standard form
\begin{multline}
 I_{\nu}^{imp}(\bp) = \sum_{\nu'}\int \frac{d^2p'}{(2\pi\hbar)^2}\,
  W_{\bp\bp'}^{\nu\nu'}\,\delta(\eps_{\nu}-\eps_{\nu'})
  \\ \times
  \left[f_{\nu'}(\bp') - f_{\nu}(\bp)\right]
 \label{Iimp-1}
\end{multline}
and
\begin{multline}
 I_{\nu}^{ee}(\bp) = \sum_{\nu_1} \sum_{\nu_2} \sum_{\nu_3}
 \int\frac{d^2p_1}{(2\pi\hbar)^2} \int\frac{d^2p_2}{(2\pi\hbar)^2} \int d^2p_3\,
 \\ \times
 \delta(\bp + \bp_1  -  \bp_2 - \bp_3)\,\delta(\eps_{\nu} + \eps_{\nu_1} - \eps_{\nu_2} - \eps_{\nu_3})
 \\ \times
 W^{\nu\nu_1,\nu_2\nu_3}_{\bp\bp_1, \bp_2 \bp_3} 
 \\ \times
 \bigl[ (1-f)(1-f_1)\,f_2\,f_3 - f\,f_1\,(1-f_2)(1 - f_3) \bigr].
 \label{Iee-1}
\end{multline}
We assume that point-like impurities with concentration $n_i$ are described by the potential $U(\br) = U_0\,\delta(\br)$,
so the electron--impurity scattering rate in the Born approximation calculated using $\psi_{\bm\nu}$ from Eq. \eqref{WF} equals
\be
  W_{\bp\bp'}^{\nu\nu'} = \frac{\pi}{\hbar}\,n_i U_0^2\,
 [1 + \nu\,\nu' \cos(\widehat{\bp,\bp'})].
 \label{W2}
\ee
We also assume that due to the screening by a nearby gate, the interaction potential is short-ranged and may be written in the 
form $V(\br-\br') = V_0\,\delta(\br-\br')$. In the Born approximation, the scattering rate is proportional to the square of the 
difference between the matrix element of direct and exchange interaction 
$|\la\bp\nu,\bp_1\nu_1|V|\bp_2\nu_2,\bp_3\nu_3\ra - \la\bp\nu,\bp_1\nu_1|V|\bp_3\nu_3,\bp_2\nu_2\ra|^2$, 
where
\begin{multline}
 \la\bp\nu,\bp_1\nu_1|V|\bp_2\nu_2,\bp_3\nu_3\ra = \sum_{\sigma} \sum_{\sigma'} \int d\br \int d\br'\,
 \psi_{\bp\nu}^{(\sigma)^*}(\br)
 \\ \times
 \psi_{\bp_1\nu_1}^{(\sigma')^*}(\br')\,V(\br-\br')\,
 \psi_{\bp_2\nu_2}^{(\sigma)}(\br)\,\psi_{\bp_3\nu_3}^{(\sigma')}(\br').
 \label{matr_el}
\end{multline}
Making use of the explicit form of $\psi_{\bp\nu}^{(\sigma)}(\br)$ Eq. \eqref{WF}, one easily obtains that 
\begin{multline}
 W^{\nu\nu_1,\nu_2\nu_3}_{\bp\bp_1, \bp_2 \bp_3} 
 = \frac{\pi}{2}\,\frac{V_0^2}{\hbar}\,
 [1 - \nu\,\nu_1 \cos(\widehat{\bp,\bp_1})]
 \\ \times
 [1 - \nu_2\,\nu_3 \cos(\widehat{\bp_2,\bp_3})].
 \label{W4}
\end{multline}

As the spectrum of the system  is rotationally symmetric, it is convenient to seek the linear response to the
electric field $\bm E$ or the temperature gradient $\nabla T$ in the form
\be
 f_{\nu}(\bp) = \bar f(\eps_{\nu}) + C_{\nu}(p)\,\bar f(\eps_{\nu})\,[1 - \bar f(\eps_{\nu})]\cos\p,
 \label{f-ansatz}
\ee
where $\bar{f}$ is the equilibrium Fermi distribution, $p$ is the absolute value of $\bp$, and $\p$ is the angle
between $\bm E$ or $\nabla T$ and $\bp$. The temperature is assumed to be low, so the nonequilibrium correction
to $\bar f$ is nonzero only near the Fermi energy. With this substitution, the linearization of Eq. \eqref{Iee-1} 
results in the replacement of the distribution-dependent factor in it by the expression \cite{Haug-book}
\begin{multline}
 \left(C_{\nu_2}\cos\p_2 + C_{\nu_3}\cos\p_3 - C_{\nu}\cos\p - C_{\nu_1}\cos\p_1 \right)
 \\ \times
 (1-\bar{f})(1-\bar{f}_1)\,\bar{f}_2\,\bar{f}_3.
 \label{C_cos}
\end{multline}
To proceed further, it is convenient to replace the integration variables $\bp_i$ in Eq. \eqref{Iee-1} by $\eps_i$ and
$\p_i$. 
%
%
This replacement is straightforward at $\eps>0$ because $p$ is a single-valued function of energy for both spectrum
branches, but at $\eps<0$ there is only one branch $\nu=-1$ and any value of $\eps$ corresponds to two values of $p$ 
(see Fig.~\ref{fig:bands}).
To overcome this difficulty, we replace the branch indices $\nu$ in Eq. \eqref{Boltz-1} by indices $\mu=\pm 1$ that
correspond to the smaller and larger momentum for a given $\eps$, hence $p_{\mu}(\eps)$ are single-valued functions.
A substitution of Eq. \eqref{f-ansatz} into the collision integral with impurities Eq. \eqref{Iimp-1}  gives
\begin{multline}
 I_{\mu}^{imp}(\eps,\p) = 
 -\Gamma_0\, \cos\p\,{\bar f}(1 - \bar f)\,
\\ \times
 \frac{ (p_{\mu} + 2\,p_{-\mu})\,C_{\mu} + \sgn\eps\, p_{-\mu}\,C_{-\mu} }{p_{\mu} + p_{-\mu}},
 \label{Iimp-2}
\end{multline}
where $\Gamma_0 = n_i U_0^2\,(p_{\mu} + p_{-\mu})/4\hbar^3 |v_{\mu}|$ and $v_{\mu}^{-1}= \partial p_{\mu}/\partial\eps$.
Defined in this way, $\Gamma_0$ exhibits a peculiarity at the bottom of the lower helicity band due to the singularity in the density of states but is constant at high energies.

The calculation of the electron--electron collision integral is much more involved. Assuming that all the quantities except
the distribution functions are energy-independent near the Fermi level and calculating the phase volume available for the 
scattering of electrons with given energies as in Ref. \cite{Nagaev08} (See Appendix  \ref{A:angular} for the details), one 
finally obtains
\begin{multline}
 I_{\mu}^{ee}(\eps,\p) = 2\cos\p\,\frac{\Gamma_2}{T^2} \int d\eps'\,K(\eps,\eps')
\\ \times
 \biggl\{
  Q_{\mu}  \left[ C_{\mu}(\eps') - C_{\mu}(\eps) \right]
 +
  \Psi_{\mu}\,
  \frac{p_{\mu}\,C_{-\mu}(\eps') - p_{-\mu}\,C_{\mu}(\eps')}{p_{\mu} + p_{-\mu}}
\\  +
  \sum_{\mu_1} R_{\mu\mu_1}\,\left[C_{\mu_1}(\eps')  - C_{\mu_1}(-\eps') \right]
 \biggr\},
 \label{I-E-1}
\end{multline}
where $\Gamma_2(T)=V_0^2 T^2\,(p_{\mu}+p_{-\mu})/32\pi^3\hbar^5|v_{\mu}|^3$ is the effective rate of electron--electron
collisions,
\be
 K(\eps,\eps') = \bigl[1 - \bar{f}(\eps)\bigr]\,\frac{\eps-\eps'}{e^{(\eps-\eps')/T} - 1}\,\bar{f}(\eps'),
 \label{K}
\ee
and all the energies are measured from $E_F$. 

\begin{figure}
 \includegraphics[width=0.9\columnwidth]{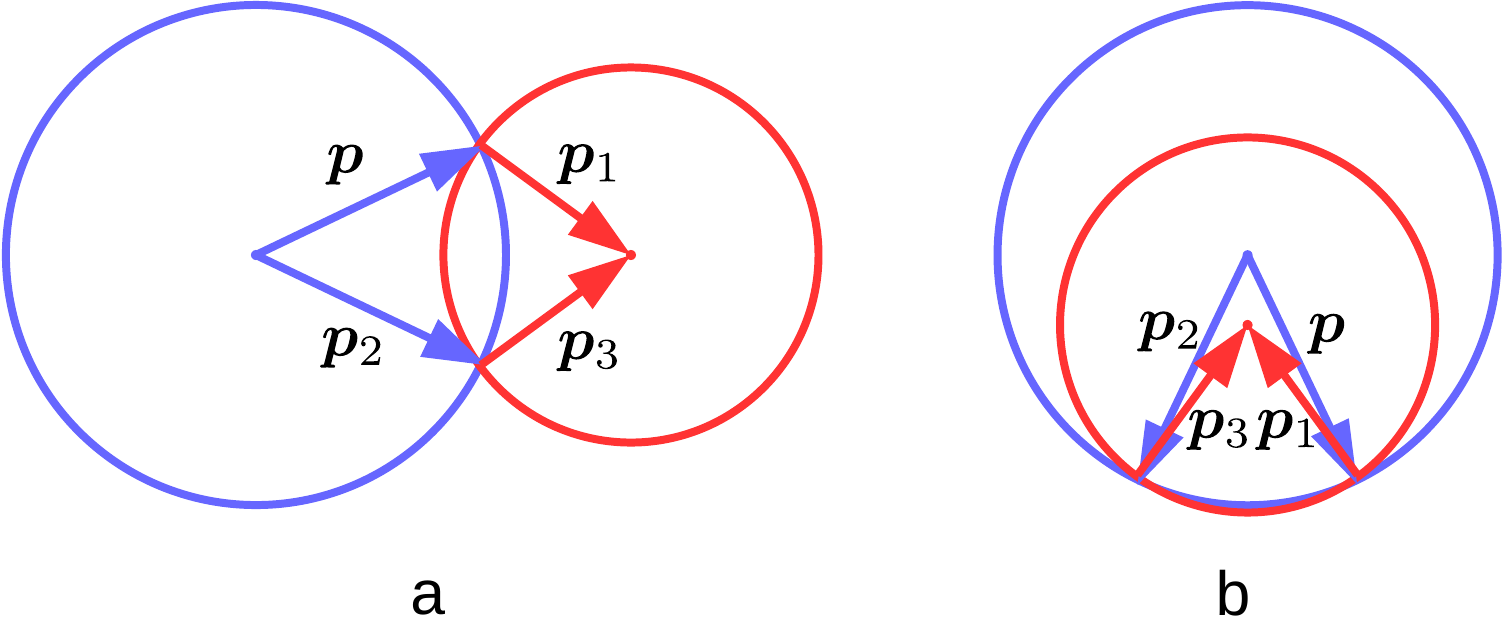}
 \caption{\label{fig:processes} The origin of the logarithmic singularity in Eq. \eqref{Q}. If an electron with momentum $\bp$
 from the outer Fermi contour collides with an electron with momentum $\bp_1$ from the inner contour and they are scattered again 
 to different contours, the states $\bp_1$ must lie at the intersections of two Fermi contours, one of which is shifted by $\bp+\bp_1$.
 If $\bp$ and $\bp_1$ are parallel or antiparallel, these contours become externally or internally tangent rather than intersecting, 
 and hence the number of states participating in the scattering sharply increases.}
\end{figure}

The first term in Eq. \eqref{I-E-1} is similar to the expression that arises in 2D conductors with a singly
connected Fermi surface. The coefficient $Q_{\mu}$ diverges at $T\to 0$ and its most singular part is of the form
\be
 Q_{\mu} = 4\,\frac{p_{\mu} + 3\,p_{-\mu}}{p_{\mu} + p_{-\mu}}\,\ln\frac{E_F}{T},
 \label{Q}
\ee
but this term vanishes for $C_{\mu}(\eps)={\rm const}$ and does not affect the electric resistivity if taken alone. The
logarithmic singularity in the scattering rate of 2D electrons with singly connected Fermi surfaces is known to result from 
their head-on or small-angle collisions \cite{Hodges71,Giuliani82}. In the case of doubly connected Fermi surface, the
singularity in Eq. \eqref{Q} arises not only from the scattering processes within the same Fermi contour, but also
from  the processes in which two pairs of the involved states, 
initial or final, belong to the same Fermi contours (see Fig. \ref{fig:processes}). In this figure, available for the
scattering states are located at the intersections of two Fermi contours, one of which is shifted by the total momentum 
of colliding electrons. It is clearly seen that when their momenta are aligned or oppositely directed so that $\p-\p_1=0$
or $\p-\p_1=\pi$, these contours become tangent rather than intersecting, hence the phase space available for scattering 
sharply increases. Depending on the specific indices $\nu\ldots\nu_3$, one of these singularities is suppressed by the 
angle-dependent factors in Eq. \eqref{W4}.

The third term in Eq. \eqref{I-E-1} also presents an extension of a similar contribution for a singly connected Fermi surface 
and contains low-temperature logarithmic singularities of the form
\begin{equation}
 R_{\mu\mu_1} = 8 \bigl[ \sgn E_F - 2\,\delta_{\mu\mu_1} \Theta(E_F) \bigr]\, 
                \frac{p_{\mu_1}}{p_{\mu} + p_{\mu_1}}\,\ln\frac{E_F}{T},
\label{R}
\end{equation}
but is zero for any even $C_{\mu}(\eps)$. This term does not contribute to the electric conductivity but is essential when dealing
with thermal transport.

\begin{figure}
 \includegraphics[width=0.9\columnwidth]{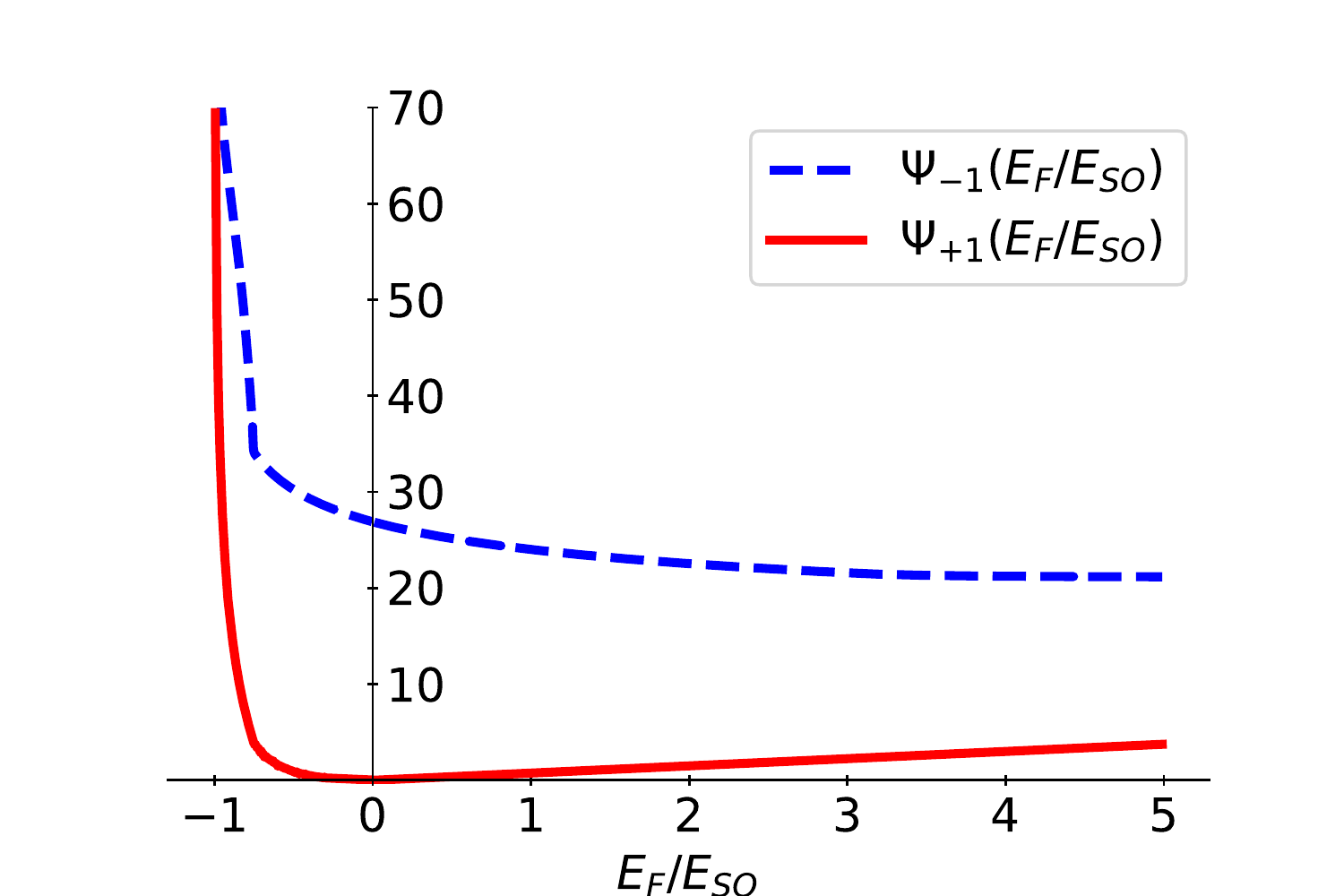}
 \caption{\label{fig:Psi} The dependences of $\Psi_{\mu}$ on $E_F/E_{SO}$.}
\end{figure}

The second term in Eq. \eqref{I-E-1} has no analog for a singly connected Fermi surface and is of special interest because it
does not vanish for arbitrary energy-independent $C_{\mu}$. The specific form of this term is best understood by comparing the
first factor in Eq. \eqref{C_cos} with the argument of the momentum delta function in Eq. \eqref{Iee-1}. As this argument must be
zero for all collisions that satisfy momentum conservation, its projection on the direction of $\bm E$ or $\nabla T$ immediately 
gives 
$$p_{\mu_2}\cos\p_2 + p_{\mu_3}\cos\p_3 - p_{\mu}\cos\p - p_{\mu_1}\cos\p_1 =0.$$ 
Therefore the first factor in Eq. 
\eqref{C_cos}  turns into zero if $C_{\mu}/p_{\mu} = C_{\mu_1}/p_{\mu_1} = C_{\mu_2}/p_{\mu_2} = C_{\mu_3}/p_{\mu_3}$ and hence
the resulting expression is proportional to $p_{\mu}\,C_{-\mu}(\eps') - p_{-\mu}\,C_{\mu}(\eps')$. The factors $\Psi_{\mu}$ are
given by integrals that can be calculated only numerically (see Appendix \ref{A:angular}). The $\Psi_{\mu}(E_F)$ curves are shown 
in Fig. \ref{fig:Psi}. Both of them exhibit a logarithmic singularity at the bottom of the lower helicity band $E_F=-S_{SO}$ and tend to
the same value $\Psi_{\pm1}=16$ at $E_F \gg E_{SO}$. However $\Psi_1$ monotonically decreases with increasing $E_F$, while
$\Psi_{-1}$ first decreases to zero at $E_F=0$ and then increases again. 
{ The kinks in $\Psi_{\mu}$ at $E_F<0$ is due to the closure of scattering channels with three electron states
at the inner Fermi contour and one at the outer contour at $p_1=3\,p_{-1}$.}
Note that unlike $Q_{\mu}$ and $R_{\mu\mu_1}$, 
$\Psi_{\mu}$ do not have a low-temperature logarithmic singularity. This is because  at $\p-\p_1=0$ and $\p-\p_1=\pi$, the 
quadrangles in Fig. \ref{fig:processes} collapse into  segments and the first factor in Eq. \eqref{C_cos} turns into zero 
regardless of the ratio of $C_1$ to $C_{-1}$. This term appears to be of crucial importance in calculating the electrical
conductivity of 2D electron gas.

\section{Electrical conductivity}\label{sec:electric}

In the linear approximation in the electric field, the Boltzmann equation Eq. \eqref{Boltz-1} assumes the form
\be
 eEv_{\mu}\,\cos\p\,\frac{\partial\bar{f}}{\partial\eps} =  I_{\mu}^{imp}(\eps,\p) + I_{\mu}^{ee}(\eps,\p),
 \label{Boltz-el-1}
\ee
where $I_{\mu}^{imp}$ and $I_{\mu}^{ee}$ are given by Eqs. \eqref{Iimp-2} and \eqref{I-E-1}.
As the perturbation in left-hand side is an even function of $\eps$ and both collision integrals conserve parity,
the solutions for $C_{\mu}$ are also even in $\eps$ and the last term in Eq.~\eqref{I-E-1} may be discarded. To solve
the system of resulting integral equations, we use the method pioneered in \cite{Brooker68} and introduce
new variables 
\be
\rho_{\mu}(\eps) = [\bar{f}\,(1-\bar{f})]^{1/2}\,C_{\mu}(\eps).
\label{rho}
\ee
%
As a result, the kernel $K(\eps,\eps')$ of the integral in Eq. \eqref{I-E-1} is replaced by a function of $\eps'-\eps$,
and the integral equations \eqref{Boltz-el-1} may be brought to the differential form by the Fourier transform
\be
 \tilde\rho_{\mu}(u) = \int d\eps\,e^{-i\eps u}\,\rho_{\mu}(\eps).
 \label{Fourier}
\ee
Furthermore, an introduction of the new independent variable $\xi=\tanh(\pi Tu)$ brings these equations to the form
\begin{multline}
 \Gamma_2 \left[
  Q_{\mu} \left(\hat{L} + 2\right)\,\tilde\rho_{\mu}
 -
 2\,\Psi_{\mu}\,\frac{p_{-\mu}\,\tilde\rho_{\mu} - p_{\mu}\,\tilde\rho_{-\mu}}
                                {p_{\mu} + p_{-\mu}}
 \right]
\\ -
 \frac{1}{\pi^2}\,
 \frac{\Gamma_0}{1-\xi^2}\,
 \frac{(p_{\mu} + 2p_{-\mu})\,\tilde\rho_{\mu} + \sgn E_F\, p_{-\mu}\,\tilde\rho_{-\mu}}
      {p_{\mu} + p_{-\mu}}
\\ =
 -\pi^{-1}{eEv_{\mu}}\,(1 - \xi^2)^{-1/2},     
 \label{Boltz-xi-el}                            
\end{multline}
where $\hat L$ stands for the differential operator
\be
 \hat{L}\,\phi = \frac{\partial}{\partial\xi}\biggl[(1-\xi^2)\,\frac{\partial\phi}{\partial\xi}\biggr]
 - \frac{\phi}{1-\xi^2}.
 \label{L}
\ee
The eigenfunctions of this operator involve Jacobi polynomials and are proportional \cite{landau-book} to $(1-\xi^2)^{1/2}\,P_m^{(1,1)}(\xi)$
 with the corresponding eigenvalues  $-(m+1)(m+2)$. As $\tilde\rho_{\mu}$ are even functions of $\xi$,
it is convenient to present them as series expansions over the normalized even-number eigenfunctions $\phi_m$ of operator
$\hat L$
\be
 \tilde\rho_{\mu}(\xi) = \sum_{m=0}^{\infty} \gamma_{\mu m}\,\phi_{2m}(\xi).
 \label{series-even}
\ee
A substitution of these expansions into Eqs. \eqref{Boltz-xi-el} and their projection on the same set of functions
results in an infinite system of equations
\begin{multline}
 2\,\Gamma_2 \biggl[
  m\,(2m+3)\,Q_{\mu}\,\gamma_{\mu m} 
  + 
  \Psi_{\mu}\,\frac{p_{-\mu}\,\gamma_{\mu m} - p_{\mu}\,\gamma_{-\mu m}}{p_{\mu} + p_{-\mu}}
 \Biggr]
 \\+
 \frac{\Gamma_0}{\pi^2} \sum_{n=0}^{\infty} Y_{mn}\,
 \frac{(p_{\mu} + 2p_{-\mu})\,\gamma_{\mu n} + \sgn E_F\, p_{-\mu}\,\gamma_{-\mu n}}{p_{\mu} + p_{-\mu}}
\\ =
 \pi^{-1}\,{eEv_{\mu}}\,X_m,
 \label{Boltz-psi}
\end{multline}
where $X_m$ and $Y_{mn}$ depend only on $m$ and $n$ with explicit expressions given in Appendix \ref{A:XYZ}.
Once the quantities $\gamma_{\mu m}$ are known, the distribution functions $f_{\mu}$ may be restored using Eqs. \eqref{series-even}, \eqref{Fourier}, \eqref{rho}, and \eqref{f-ansatz}, which results in the density of electric current of the form
\be
 j = \frac{e}{8\pi^2\hbar^2} \sum_{\mu} p_{\mu} \sgn v_{\mu}
 \sum_m X_m\,\gamma_{\mu m}.
 \label{j-series}
\ee

First consider the case where only the impurity scattering is present. The solution of Eq. \eqref{Boltz-psi} is 
\be
 \tilde\rho_{\mu}^{imp}(\xi) = \frac{2\sqrt{3}\pi}{3}\,\frac{eEv_{\mu}}{\Gamma_0}\,\frac{p_{\mu}}{p_{\mu}+p_{-\mu}}\,\phi_0(\xi),
 \label{rho_imp}
\ee
which results in the current density
\be
  j^{imp} = \frac{1}{4\pi}\,E\,\frac{e^2}{\hbar^2}\,\frac{v_{1}}{\Gamma_0}\,
  \frac{p_{1}^2 + p_{-1}^2}{p_{1} + p_{-1}}
  \label{j-imp}
 \ee
equivalent to the one obtained in Refs. \cite{Brosco16,Hutchinson18}.

As $v_{\mu}$ are equal and positive for both Fermi contours at $E_F>0$, the solution Eq. \eqref{rho_imp} also turns into 
zero the term resulting from electron--electron collisions, so they do not affect the resistivity. This is due to the specific 
form of the electron distribution Eq. \eqref{rho_imp}, which results from the particular probability of impurity scattering Eq. 
\eqref{W2} and hence from the assumption of short-ranged impurity potential. Therefore this is not a universal property of 
2D SOC electron systems (see Appendix \ref{A:impurity}).

Below the band-crossing point, $v_{\mu}$ are equal in magnitude but are of opposite signs at both Fermi contours, hence the 
distribution Eq. \eqref{rho_imp} does not turn $I^{ee}_{\mu}$ into zero and the electron--electron scattering is essential.
First we calculate the correction to the current from electron--electron collisions treating them as a perturbation in the case
of a strong impurity scattering. This is conveniently done by means of Eq. \eqref{Boltz-xi-el} as the zero-approximation
distribution \eqref{rho_imp} eliminates in it the term proportional to $Q_{\mu}$ that contains a differential operator.
Therefore the solution for the first-order correction is straightforward, and one obtains the corrections to 
$\tilde\rho_{\mu}^{imp}(\xi)$ proportional to $(1-\xi^2)^{3/2}$. The correction to the current
\be
 \delta j^{ee}
 =
 -\frac{2\pi e^2E}{3\,\hbar^2}\,
 \frac{\Gamma_2}{\Gamma_0^2}\,v_{1}\,
 \frac{p_{1}\,p_{-1} \left[ p_{-1}^2\,\Psi_{-1} + p_{1}^2\,\Psi_{1}\right]}
 {(p_{1}+p_{-1})^3}.
 \label{dj}
\ee
is proportional to $T^2$ in agreement with \cite{Pal12-LJP} and does not contain a logarithm of $T$ as one might expect for a 2D 
system. Quite predictably, it tends to zero at the band-crossing point where the inner Fermi contour shrinks to a point and
$p_{-1}=0$. It diverges as the Fermi level approaches the bottom of the lower helicity band due to the singularity in the density 
of states, but it only means that the perturbative approach fails there.

\begin{figure}
 \includegraphics[width=0.9\columnwidth]{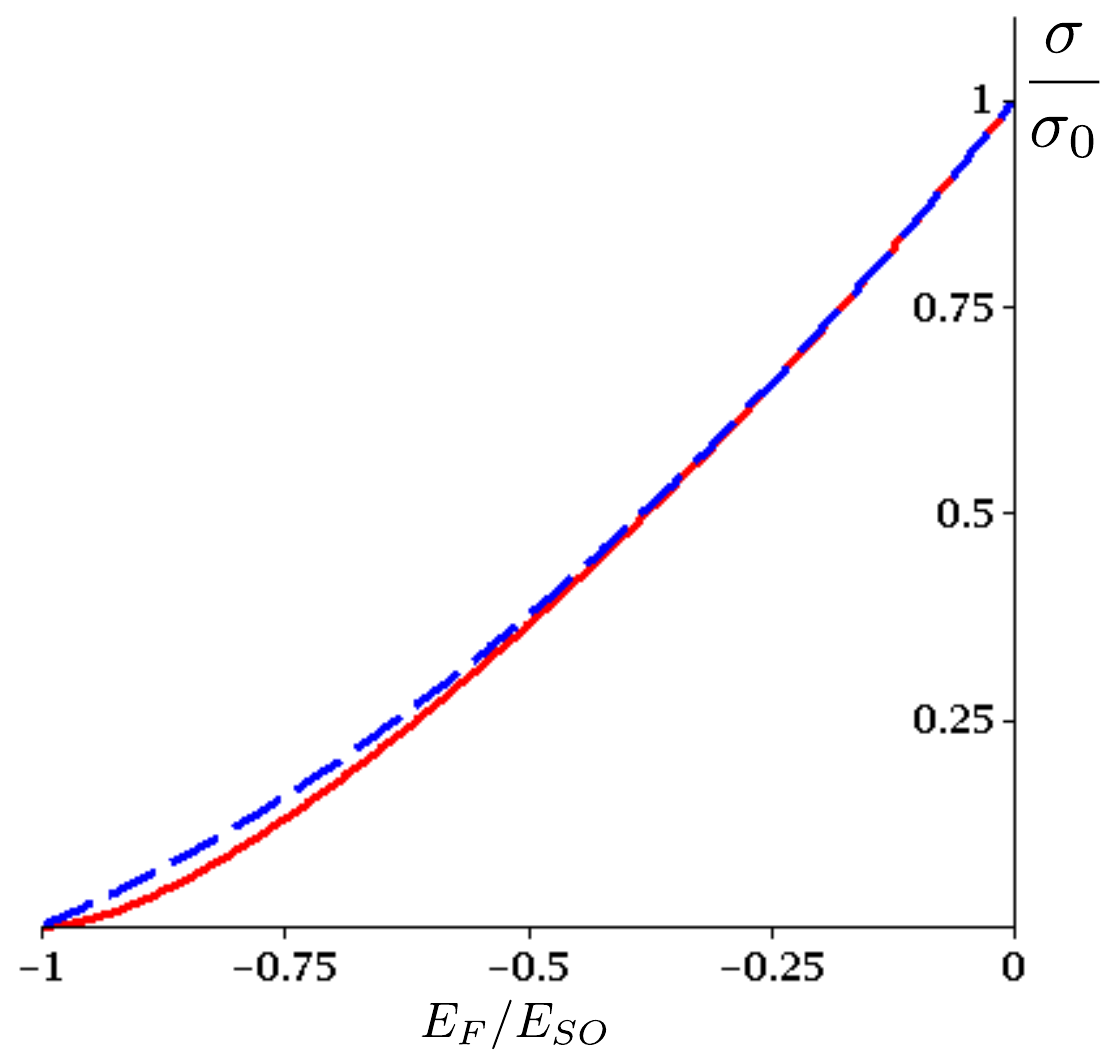}
 \caption{\label{fig:limiting} The dependences of conductivity $\sigma$ on $E_F$ in the case of strong (solid line) and  weak
 (dashed line) electron--electron scattering for the same impurity-scattering rate given by Eqs. \eqref{j-weak} and
 \eqref{j-imp}. The conductivity is normalized to its value $\sigma_0 = e^2\hbar\,\alpha^2/\pi n_i\,U_0^2$ at $E_F=0$. 
 At $E_F>0$, the electron--electron scattering has no effect on $\sigma$.}
\end{figure}

\begin{figure}
 \includegraphics[width=0.9\columnwidth]{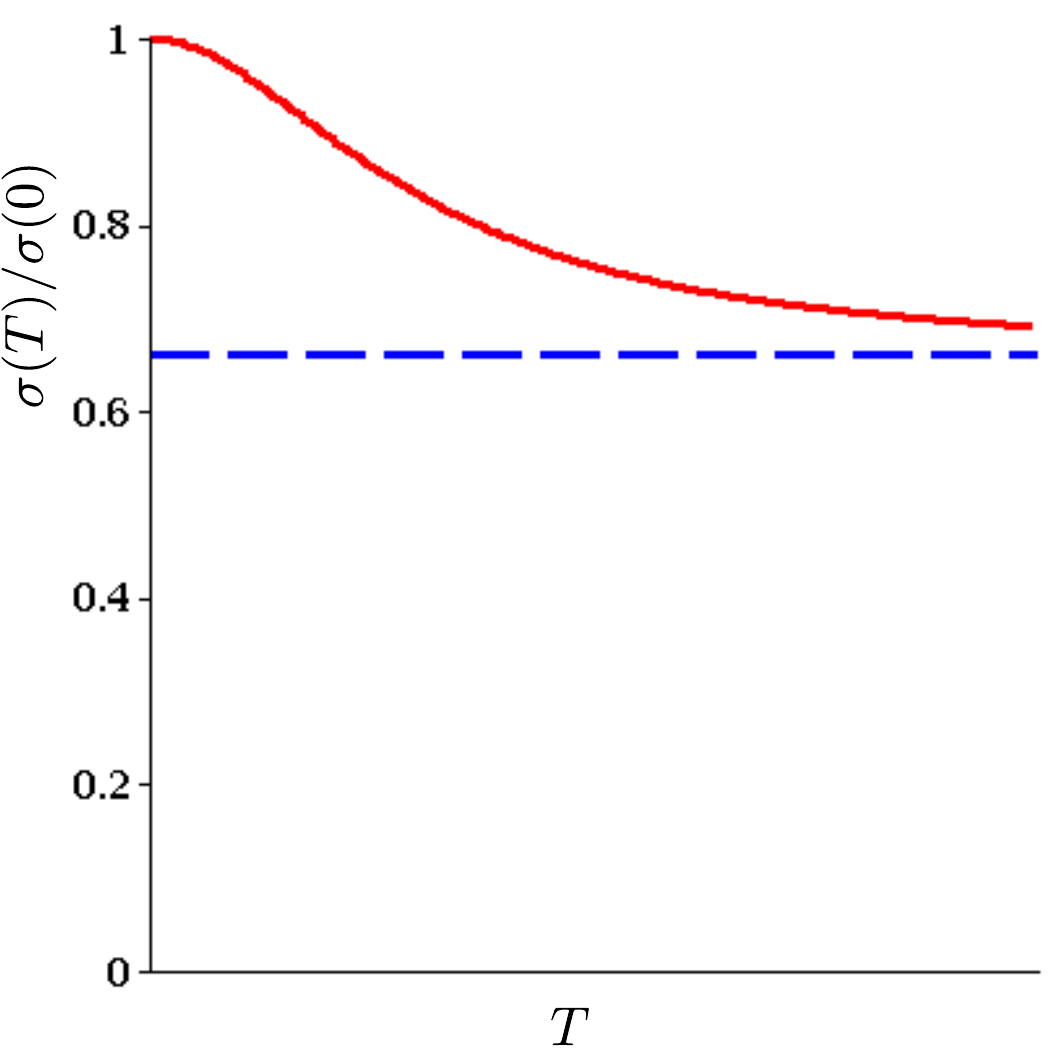}
 \caption{\label{fig:j_vs_T} The temperature dependence of normalized conductivity for $E_F = -0.85\,E_{SO}$
 obtained by a numerical solution of Eqs. \eqref{Boltz-psi}.}
\end{figure}

Consider now the opposite case of strong electron--electron scattering. It is easily seen that if one simply sets $\Gamma_0=0$,
the system of equations \eqref{Boltz-psi} for $m=0$ becomes degenerate because its left-hand side is made zero by any distribution 
with $\gamma_{10}/\gamma_{-10} = p_1/p_{-1}$. To avoid this, one has to introduce in Eqs. \eqref{Boltz-psi} a very weak impurity 
scattering. It results only in corrections of the order $1/\Gamma_2$ for $\gamma_{\mu m}$ with $m\ne 0$, but the leading terms
in $\gamma_{\mu 0}$ appear to be proportional to $1/\Gamma_0$. The leading contribution to the current density equals
\begin{multline}
 j^{ee} = E\,\frac{e^2}{4\pi\hbar}\,\frac{v_1\,(p_{1} + p_{-1})}{\hbar\Gamma_0}
\\ \times
 \frac{(p_{1}^2 - p_{-1}^2)(\Psi_{-1} - \Psi_{1})}
      {(p_{1}^2 - p_{-1}^2)(\Psi_{-1} - \Psi_{1}) + 2\,p_{1}\,p_{-1}\,(\Psi_{1} + \Psi_{-1})}.
 \label{j-weak}
\end{multline}
Though this current density is inversely proportional to the impurity-scattering rate like Eq. \eqref{j-imp}, it is 
somewhat smaller and has a different dependence on $E_F$ (see Fig. \ref{fig:limiting}). The ratio $j^{imp}/j^{ee}$
reaches its minimum value $\sim0.66$ at $E_F/E_{SO}\approx-0.85$. The two curves merge at $E_F = -E_{SO}$ and $E_F=0$.
The temperature dependence of conductivity may be obtained by truncating the infinite series Eq. \eqref{series-even} to
a finite number of terms and numerically solving the system \eqref{Boltz-psi}. The resulting curve is shown in Fig.
\ref{fig:j_vs_T} for $E_F = -0.85 E_{SO}$ and exhibits a saturation of conductivity with increasing temperature.

\section{Thermal conductivity}\label{sec:thermal}

\begin{figure}
 \includegraphics[width=0.9\columnwidth]{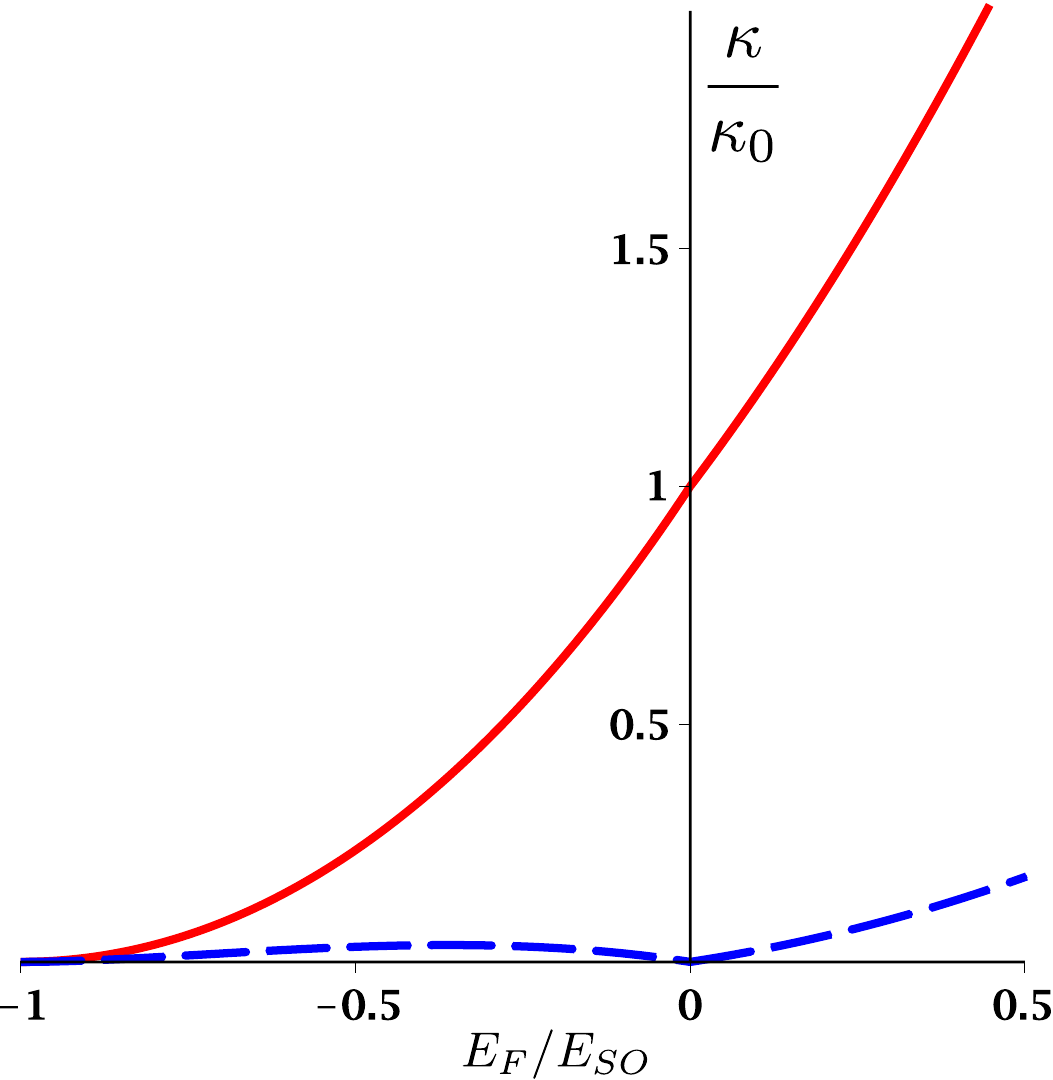}
 \caption{\label{fig:thermal} The dependence of the thermal conductivity $\kappa$ on the Fermi energy given by Eq. \eqref{k}. 
 The dashed line shows the contribution from the inner Fermi contour. The thermal conductivity is normalized to its value
 $\kappa_0 \approx 2.77\,\hbar^3\alpha^4/V_0^2 T$ at $E_F=0$.}
\end{figure}

As electron--electron collisions do not conserve heat flux, they generally limit thermal conductivity even in the
absence of additional scattering mechanisms, so there is no need to include an additional impurity scattering.
If the perturbation is caused by a gradient of temperature, the equation for the linear response is of the form
\be
 v_{\mu}\,|\nabla T|\,\frac{\eps}{T^2}\,\bar{f}\,(1-\bar{f})\,\cos\p = I_{\mu}^{ee}
 \label{Boltz-th-1}
\ee
We seek again the distribution function of electrons in the form \eqref{f-ansatz}, but now $C_{\mu}$ are odd functions of
$\eps$ according to the symmetry of perturbation. For this reason, the last term in Eq. \eqref{I-E-1} does not vanish, but
instead the second term may be omitted because it does not contain the $\ln(E_F/T)$ factor. Hence
\begin{multline}
 I^{ee}_{\mu} = -8\,\cos\p\,\frac{\Gamma_2}{T^2}\,\ln\frac{E_F}{T} \int d\eps'\,K(\eps,\eps')
 \\ \times
 \Bigl[ 
  (p_{\mu} + 3\,p_{-\mu})\,C_{\mu}(\eps) + (p_{\mu} - p_{-\mu})\,C_{\mu}(\eps')
\\ - 4\,\sgn E_F\,p_{-\mu}\,C_{-\mu}(\eps')
 \Bigr]
 (p_{\mu} + p_{-\mu})^{-1}.
 \label{I-therm}
\end{multline}
By repeating the steps described by Eqs. \eqref{rho} and \eqref{Fourier} in the previous section, one arrives at the 
equation 
\begin{multline}
 \left[ (p_{\mu} + 3\,p_{-\mu})\,\hat L - 2\,(p_{\mu} - p_{-\mu}) + 8\,\sgn E_F\,p_{-\mu} \right]
 \tilde\rho_{\mu}
\\ 
 =
 -\frac{i}{4}\,\frac{|\nabla T|\,v_{\mu}\,(p_{\mu} + p_{-\mu})}{\Gamma_2\,\ln(E_F/T)}\,
 \frac{\xi}{\sqrt{1-\xi^2}}.
 \label{Boltz-xi-th}
\end{multline}
%
%
As the right-hand side of this equation is an odd function of $\xi$, this equation can be solved by expanding
$\tilde\rho_{\mu}$ over odd-number eigenfunctions $\phi_{2m+1}$ of operator $\hat L$. In the absence of impurity
scattering, this system becomes uncoupled for different $m$ and is easily solved. The heat flux $\bf q$ is
obtained as an infinite series over $m$, and the thermal conductivity $\kappa=q/|\nabla T|$ is of the form
\begin{multline}
 \kappa = -\frac{T\,v_1\,(p_{1}+p_{-1})}{64\pi\hbar^2\Gamma_2 \ln(E_F/T)}
 \sum_{m=0}^{\infty} \frac{4m+5}{S_m\,(S_m+1)}
\\ \times
 \frac{ (3S_m-1)(p_{1}+p_{-1})^2 - 4\,(S_m-3)\,p_{1}\,p_{-1} }
      {( 3S_m-1)(p_{1}+p_{-1})^2 + 4\,(S_m-3)\,p_{1}\,p_{-1} },
 \label{k}
\end{multline}
where $S_m=(m+1)(2m+3)$.
The temperature dependence of thermal conductivity follows the same $[T\ln(E_F/T)]^{-1}$ law as for 2D electron gas
without SOC \cite{Lyakhov03}. Its dependence on the Fermi level is shown in Fig. \ref{fig:thermal}. Though the  
expression for $\kappa$ \eqref{k} does not explicitly depend on the sign of $E_F$, it exhibits a kink at $E_F=0$ because 
the derivative of the smaller Fermi momentum $dp_{-1}/dE_F$ changes its sign at this point.The relative change of the
slope is 
\be
 \frac{d\kappa/dE_F|_{+0}}{d\kappa/dE_F|_{-0}} \approx 0.83
 \label{jump}
\ee
This relation is free from any unknown parameters and can serve as an experimental test of the considered model. The
negative jump of the derivative $d\kappa/dE_F$ results from the peculiarity in the scattering of electrons on the outer Fermi 
contour by the electrons on the inner Fermi contour at the band-crossing point. On the contrary, the heat flux carried by the
electrons on the inner contour turns into zero at this point and therefore exhibits a positive jump of derivative.

\section{Discussion}\label{sec:discussion}

The effects considered in previous sections are best observed in 2D electron systems with strong SOC like InAS, which exhibits 
Rashba parameter $\hbar\alpha =1.2\,{\rm eV\AA}$ \cite{Heedt2017}. The temperature should be sufficiently low to suppress the 
electron--phonon scattering, which is proportional to $T^{4.5}$ in 2D systems \cite{Kawamura92}. Furthermore, the parameter
of electron--electron scattering $\Gamma_2$ has to be larger than the impurity-scattering parameter $\Gamma_0$. At $T=2$ K, the
electron concentration $10^{10}$ cm$^{-2}$, and the gas--gate distance of 20 nm, one obtains the transport scattering length 
$l_{ee} = |v_{\mu}|/\Gamma_2 \sim 250$ nm. This is well below the elastic mean free path of 800 nm reported very recently in InAs
2D electron gas in Ref. \cite{Lee19}, so the regime of dominant electron--electron scattering may be achieved for realistic 
parameters of the system.

While the thermal transport in 2D electron systems with strong electron--electron scattering is only slightly affected by SOC, its 
effects on charge transport in these systems are much less trivial. In the absence of SOC, this type of scattering does not affect 
charge transport at all because of momentum conservation. One may think that the emergence of a double Fermi contour will lift 
this constraint  and the electron--electron collisions will become the dominant mechanism of current relaxation, but this is not 
the case. The reason is that a certain type of perturbation involving the electron distributions on both contours is not affected 
by them. As a result, increasing the intensity of electron--electron scattering does not fully suppress the current induced by 
applied electric field \cite{*[{For single-band metals, it was noted in }] [{}] Maslov11}. Instead, this current decreases only to 
a finite value, which is determined by other mechanisms of scattering and depends on their details as well as  those of 
electron--electron interaction. Though we made explicit calculations for a point-like interaction potential, these conclusions are 
qualitatively valid for its arbitrary shape.

The partial nature of current relaxation via electron--electron collisions is not unique to 2D systems with SOC. The existence of 
the perturbation immune to electron--electron collisions is a general property of systems with multiply connected Fermi surface, 
which results from the momentum and energy conservation and the structure of the electron--electron collision integral.
This perturbation is unaffected even by triple electronic collisions \cite{Lunde07} because they obey the same conservation laws.
Therefore a similar partial relaxation of the current by these collisions may be observed in a broad class of 2D and 3D systems. 
Possible candidates are graphene with Zeeman-shifted Dirac points or 2D systems without SOC but with two filled 
transverse subbands.

\begin{acknowledgements}

This work was supported by Russian Science Foundation (Grant No 16-12-10335).

\end{acknowledgements}

\appendix

\section{Angular integration in the collision integral} \label{A:angular}

Here we present a derivation of the collision integral Eq. \eqref{I-E-1} for the case where $E_F>0$. The extension to negative $E_F$ is
straightforward. We start with Eqs. \eqref{Iee-1} and \eqref{C_cos} and step by step  eliminate the integrations over
momentum angles in them. When integrating over $\bp_2$ and $\bp_3$, it is convenient to measure them with respect to the sum
$\bp_{\Sigma} \equiv \bp + \bp_1$. It is easily seen from the cosine theorem that 
\be
 \cos(\p_2-\p_3) = \frac{p_{\Sigma}^2 - p_{2}^2 - p_{3}^2}{2\,p_{2}\,p_{3}},
 \label{cos2-3}
\ee
where
\be 
 p_{\Sigma}^2 = p_{}^2 + p_{1}^2 + 2\,p_{}\,p_{1} \cos(\p_1-\p).
 \label{p+}
\ee
For brevity, we use here the notation $p_i \equiv p_{\mu_i}(\eps_i)$. The cosines of $\p_2$ and $\p_3$ are conveniently presented
in the form
\begin{multline}
 \cos\p_{2,3} = \cos(\p_{2,3}-\p_{\Sigma})\,\cos\p_{\Sigma} 
\\
 - \sin(\p_{2,3}-\p_{\Sigma})\,\sin\p_{\Sigma}.
 \label{cos2,3}
\end{multline}
It should be noted that the terms with $\sin(\p_{2,3}-\p_{\Sigma})$ vanish upon the integration over $\bp_{2,3}$
because of the symmetry, and the corresponding cosine may be expressed through the cosine theorem as
\be
 \cos(\p_{2,3}-\p_{\Sigma}) = 
 \frac{p_{{2,3}}^2 + p_{\Sigma}^2 - p_{{3,2}}^2}
 {2\,p_{{2,3}}\,p_{\Sigma}},
 \label{cos-}
\ee
while
\be
 \cos\p_{\Sigma} = \frac{p\cos\p + p_{1}\cos\p_1}{p_{\Sigma}}.
\ee
Therefore the two last cosine-dependent factors may be put before the integrals over $\bp_2$ and $\bp_3$ . The remaining
integral has been calculated in Ref. \cite{Nagaev08} and equals
\begin{multline}
 \int\frac{d^2p_2}{(2\pi\hbar)^2} \int\frac{d^2p_3}{(2\pi\hbar)^2}\,
 \delta(\eps_{\mu_2\bp_2} - \eps_2)\,\delta(\eps_{\mu_3\bp_3} - \eps_3)
\\ \times
 \delta(\bp + \bp_1 - \bp_2 - \bp_3)
 = \frac{1}{4\pi^4\hbar^4}\,\frac{p_{2}\,p_{3}}{|v_{2}\,v_{3}|}\,
 \frac{\Theta(\Delta)}{\sqrt{\Delta}},
 \label{dp2dp3}
\end{multline}
where
\be
 \Delta = \left[p_{\Sigma}^2 - (p_{2} - p_{3})^2\right]
          \left[(p_{2} + p_{3})^2 - p_{\Sigma}^2\right].
 \label{Delta-1}
\ee
Hence the integral Eq. \eqref{Iee-1} may be brought to the form
\begin{multline}
 I_{\mu}^{ee}(\eps,\p) = \cos\p\,\frac{V_0^2}{16\pi^3\hbar^5} \sum_{\mu_1} \sum_{\mu_2} \sum_{\mu_3}
 \int d\eps_1 \int d\eps_2 
\\ \times
 \int d\eps_3\, \delta(\eps + \eps_1 - \eps_2 - \eps_3)\,
 (1 - \bar{f})(1 - \bar{f}_1)\,\bar{f}_2\,\bar{f}_3
 \\ \times
 p_{1}\,
 \int_{-\pi}^{\pi} d\chi\,(1 - \mu\,\mu_1\cos\chi)
 \left| {\rm Re}\,
 \frac{{\cal D}_{\mu..\mu_3}^{\mu_2\mu_3/2}(\chi)}{v_{1}\,v_{2}\,v_{3}}
 \right|
\\ \times
 \Bigl[
    \lambda_{\mu..\mu_3}\,C_{\mu_2}
    +
    \bar\lambda_{\mu..\mu_3}\,C_{\mu_3}
    -
    \cos\chi\,C_{\mu_1} - C_{\mu}
 \Bigr],
 \label{I-sep-2}
\end{multline}
where $\mu..\mu_3$ stands for $\mu\mu_1\mu_2\mu_3$,
\be
 {\cal D}_{\mu..\mu_3} 
 = \frac{(p_{2} + p_{3})^2 - p^2 - p_{1}^2 - 2\,p\,p_{1}\cos\chi}
        {p_{}^2 + p_{1}^2 + 2\,p\,p_{1}\cos\chi - (p_{2} - p_{3})^2},
 \label{D}
\ee

\begin{multline}
 \lambda_{\mu..\mu_3}
 = 
 \frac{1}{2}\,
 \frac{p_{2}^2 - p_{3}^2 + p_{}^2 + p_{1}^2 + 2\,p_{}\,p_{1}\cos\chi}
      {p_{}^2 + p_{1}^2 + 2\,p_{}\,p_{1}\cos\chi}
\\ \times      
 (p_{} + p_{1}\cos\chi)/{p_{2}},
\end{multline}
and
\be
 \bar\lambda_{\mu\mu_1\mu_2\mu_3} \equiv \lambda_{\mu\mu_1\mu_3\mu_2}.
  \label{lambda}
\ee
If $I_{\mu}^{ee}$ is calculated in the leading approximation to the order $T^2$, all the 
quantities except the distribution functions may be considered as energy-independent near the
Fermi level. Therefore the integration over $\eps_1$ , $\eps_2$ , and $\eps_3$ is easily performed 
and the collision integral \eqref{Iee-1} is brought to the form
\begin{multline}
 I_{\mu}^{ee}(\eps,\p) = \cos\p\,\frac{V_0^2}{16\pi^3\hbar^5} \sum_{\mu_1} \sum_{\mu_2} \sum_{\mu_3}
\\ \times
 p_{1}\,
 \int_{-\pi}^{\pi} d\chi\,(1 - \mu\,\mu_1\cos\chi)
 \left| {\rm Re}\,
 \frac{{\cal D}_{\mu..\mu_3}^{\mu_2\mu_3/2}(\chi)}{v_{1}\,v_{2}\,v_{3}}
 \right|
\\ \times
 \int d\eps'\,\Bigl\{ K(\eps,\eps')
 \bigl[\lambda_{\mu..\mu_3}\,C_{\mu_2}(\eps') + \bar\lambda_{\mu..\mu_3}\,C_{\mu_3}(\eps') 
\\   - C_{\mu}(\eps)\bigr]
 - K(\eps,-\eps')\cos\chi\,C_{\mu_1}(\eps')
 \Bigr\},
 \label{I-sep-3} 
\end{multline}
where $K(\eps,\eps')$ is given by Eq. \eqref{K}. Upon regrouping the terms in Eq. \eqref{I-sep-3}, one obtains
Eq. \eqref{I-E-1}, where
\begin{multline}
 Q{\mu} = \sum_{\mu_1} \sum_{\mu_2} \sum_{\mu_3} \frac{p_{\mu_1}}{p_{\mu} +p_{-\mu}}
 \int_{-\pi}^{\pi} d\chi\,(1 - \mu\,\mu_1\cos\chi)
\\ \times
 \Theta({\cal D}_{\mu..\mu_3})\,
 {\cal D}_{\mu..\mu_3}^{\mu_2\mu_3/2},
 \label{Q-def}
\end{multline}

\begin{multline}
 \Psi_{\mu} = \sum_{\mu_1} \sum_{\mu_2} \sum_{\mu_3} \frac{p_{\mu_1}}{p_{-\mu}}
 \int_{-\pi}^{\pi} d\chi\,(1 - \mu\,\mu_1\cos\chi)\,
 \Theta({\cal D}_{\mu..\mu_3})
\\ \times
 {\cal D}_{\mu..\mu_3}^{\mu_2\mu_3/2}\, 
 \left( 1 - \delta_{\mu\mu_1}\cos\chi - 2\,\delta_{\mu\mu_2}\lambda_{\mu..\mu_3} \right),
 \label{Psi-def}
\end{multline}
and
\begin{multline}
 R_{\mu\mu_1} = \sum_{\mu_2} \sum_{\mu_3} \frac{p_{\mu_1}}{p_{\mu} +p_{-\mu}}
 \int_{-\pi}^{\pi} d\chi\,(1 - \mu\,\mu_1\cos\chi)
\\ \times
 \Theta({\cal D}_{\mu..\mu_3})\,
 {\cal D}_{\mu..\mu_3}^{\mu_2\mu_3/2}\,
 \cos\chi.
 \label{R-def}
\end{multline}
For negative $E_F$ , these quantities are given by similar expressions except that the prefactors 
$\mu\mu_1$  to $\cos\chi$ and the products $\mu_2\mu_3$ in the exponents in Eqs. \eqref{Q-def} - \eqref{R-def}
are replaced by 1.

\section{Expressions for eigenfunctions and expansion coefficients} \label{A:XYZ}

The normalized eigenfunctions of differential operator $\hat L$ defined in Eq. \eqref{L} are given by equation
\be
 \phi_m(\xi) = \sqrt{\frac{(2m+3)(m+2)}{8\,(m+1)}} \sqrt{1-\xi^2}\,P_m^{(1,1)}(\xi),
 \label{psi_m}
\ee
where $P_m^{(1,1)}(\xi)$ are Jacobi polynomials. The quantity $(1-\xi^2)^{-1/2}$ in the right-hand side of
Eq. \eqref{Boltz-xi-el} may be presented as a series
\be
 \frac{1}{\sqrt{1-\xi^2}} = \sum_{m=0}^{\infty} X_m\,\phi_{2m}(\xi),
 \label{sqrt-series}
\ee
where 
\be
 X_m = \int_{-1}^1 d\xi\, \frac{\phi_{2m}(\xi)}{\sqrt{1-\xi^2}}
 =
 \sqrt{\frac{4m+3}{(2m+1)(m+1)}}.
 \label{X}
\ee
The matrix elements of $1/(1-\xi^2)$ between the eigenfunctions of $\hat L$ are given by the equation
\begin{multline}
 Y_{mn} = \int_{-1}^1 d\xi\,\frac{\phi_{2m}(\xi)\,\phi_{2n}(\xi)}{1 - \xi^2}
 =
 \frac{\min(m,n) + 1/2}{\max(m,n) + 1}
\\ \times
 \sqrt{\frac{ (4m+3)(m+1)(4n+3)(n+1) }{ (2m+1)(2n+1) }}.
 \label{Y}
\end{multline}

\section{Momentum-dependent impurity scattering} \label{A:impurity}

If the impurities are rotationally symmetric but of finite size, the matrix element of electron--impurity 
interaction depends on the change of electron momentum $\bp-\bp'$ and hence Eq. \eqref{W2} assumes the form
\be
 W_{\bp\bp'}^{\nu\nu'} = \frac{\pi}{\hbar}\,n_i\,
 \bigl|U(p,p',\p-\p')\bigr|^2\,[1 + \nu\nu'\cos(\p-\p')\bigr],
 \label{W2pp'}
\ee
where
\be
 U(p,p',\p-\p') = \int d^2r\, e^{i(\bp'-\bp)\br/\hbar}\,U(r).
 \label{Upp'}
\ee
To be definite, we consider the case of positive $E_F$. Using the ansatz \eqref{f-ansatz} for the distribution 
function, one obtains the electron--impurity collision integral in the form
\begin{multline}
 I_{\mu}^{imp}(\eps,\p) = -\cos\p\, \bar{f}\,(1-\bar{f})
\\ \times
 \frac{(p_{\mu}\,\Gamma_{0\mu} + p_{-\mu}\,\Gamma_0')\,C_{\mu} - p_{-\mu}\,\Gamma_0''\,C_{-\mu}}
      {p_{\mu} + p_{-\mu}},
 \label{Iimp-pp'}
\end{multline}
where
\begin{multline}
 \Gamma_{0\mu} = \frac{\pi}{\hbar}\,\frac{n_i\,(p_{\mu}+p_{-\mu})}{(2\pi\hbar)^2 v}
 \int_{-\pi}^{\pi} d\chi\,(1-\cos^2\chi)
\\ \times 
 |U(p_{\mu},p_{\mu},\chi)|^2,
 \label{Gamma_mu}
\end{multline}

\begin{multline}
 \Gamma_0' = \frac{\pi}{\hbar}\,\frac{n_i\,(p_{\mu}+p_{-\mu})}{(2\pi\hbar)^2 v}
 \int_{-\pi}^{\pi} d\chi\,(1-\cos\chi)
\\ \times 
 |U(p_{\mu},p_{-\mu},\chi)|^2,
 \label{Gamma'}
\end{multline}
and
\begin{multline}
 \Gamma_0'' = \frac{\pi}{\hbar}\,\frac{n_i\,(p_{\mu}+p_{-\mu})}{(2\pi\hbar)^2 v}
 \int_{-\pi}^{\pi} d\chi\,\cos\chi\,(1-\cos\chi)
\\ \times 
 |U(p_{\mu},p_{-\mu},\chi)|^2.
 \label{Gamma''}
\end{multline}
By solving Eq. \eqref{Boltz-el-1} with $I_{\mu}^{ee}=0$, one immediately obtains the ratio
\be
 \frac{C_{\mu}^{imp}}{C_{-\mu}^{imp}}
 =
 \frac{ p_{\mu}\, \Gamma_0' + p_{-\mu}\,(\Gamma_{0,-\mu}+\Gamma'')}
      { p_{-\mu}\,\Gamma_0' + p_{\mu}\, (\Gamma_{0\mu}  +\Gamma'')}.
 \label{C/C}
\ee
This suggests that in general, $C_{\mu}^{imp}/C_{-\mu}^{imp} \ne  p_{\mu}/p_{-\mu}$,  and
the corresponding distribution function does not turn the electron--electron collision integral into zero.

\bibliography{SOI-2D,ee,books}

\end{document}